\documentclass[paper=a4,DIV=12,numbers=noendperiod]{scrartcl}

\pdfoutput=1

\usepackage{xparse}  
\usepackage{xspace}  

\usepackage[utf8]{inputenc}
\DeclareUnicodeCharacter{03BC}{\ensuremath{\mu}}
\usepackage[T1]{fontenc}
\usepackage[UKenglish]{babel}
\renewcommand{\dateUKenglish}{\def\today{\number\day~%
  \ifcase \month \or January\or February\or March\or April\or May\or June\or July\or August\or September\or October\or November\or December\fi\space%
  \number\year}}
\usepackage{csquotes}
\usepackage[babel,final]{microtype}
\usepackage{lmodern}
\usepackage[scaled=0.92]{helvet}

\usepackage{setspace}

\usepackage{scrpage2}  
\clearscrheadings
\setkomafont{pageheadfoot}{\normalfont}
\setkomafont{pagenumber}{\normalfont}

\addtokomafont{section}{\boldmath}  
\addtokomafont{subsection}{\boldmath}
\addtokomafont{paragraph}{\boldmath}  
\addtokomafont{subparagraph}{\boldmath}

\deffootnote[1em]{1em}{1em}{\textsuperscript{\thefootnotemark}\hspace{0.1em}}

\usepackage[subfigure]{tocloft}  
\newcounter{lofdepth}
\newcounter{lotdepth}
\usepackage[fleqn]{amsmath}
\usepackage{amsfonts}
\usepackage{amssymb}
\usepackage{amsthm}
\usepackage{mathtools}
\numberwithin{equation}{section}  
\usepackage{units}  
\usepackage{bm}  
\usepackage{tensor}  
\usepackage{bbm}  
\usepackage{braket}  

\usepackage{graphicx}
\usepackage{xcolor}
\definecolor{TUM}{cmyk}{1,0.43,0,0}
\definecolor{blau1}{cmyk}{1,0.57,0.12,0.7}
\definecolor{blau2}{cmyk}{1,0.54,0.04,0.19}
\definecolor{blau3}{cmyk}{0.9,0.48,0,0}
\definecolor{blau4}{cmyk}{0.65,0.19,0.01,0.04}
\definecolor{blau5}{cmyk}{0.42,0.09,0,0}
\usepackage[format=plain,font=small,labelfont=bf,width=0.85\textwidth]{caption}  
\usepackage[list=true,labelformat=simple]{subcaption}  
\usepackage{enumitem}
\usepackage{booktabs}
\usepackage{tabu}
\usepackage{tikz}
\usetikzlibrary{backgrounds,positioning,arrows,shapes.geometric,external}

\usepackage[backend=biber,style=numeric-comp,sorting=none,firstinits=true,uniquename=mininit,maxnames=15,maxcitenames=2,minnames=1,doi=false,url=false]{biblatex}
\AtEveryBibitem{%
  \clearfield{month}%
  \clearfield{issn}%
  \clearfield{subtitle}%
  \clearfield{note}%
}

\usepackage[colorlinks=true,linkcolor=blau2,citecolor=blau2,urlcolor=blau3,unicode]{hyperref}
\addto\extrasUKenglish{%
}

\newcommand*{\rep}[2][]{\ensuremath{{\bm{#2}#1}}}  
\newcommand*{\crep}[2][]{\ensuremath{{\overline{\bm{#2}}#1}}} 

\newcommand*{\SU}[1]{\ensuremath{\mathrm{SU}(#1)}}  
\newcommand*{\UU}[1]{\ensuremath{\mathrm{U}(#1)}}  
\newcommand*{\SO}[1]{\ensuremath{\mathrm{SO}(#1)}}  
\newcommand*{\OO}[1]{\ensuremath{\mathrm{O}(#1)}}  
\newcommand*{\Sp}[1]{\ensuremath{\mathrm{Sp}(2#1)}}  
\newcommand*{\USp}[1]{\ensuremath{\mathrm{USp}(2#1)}}  

\newcommand*{\lalg}[1]{\ensuremath{\mathfrak{#1}}}  

\newcommand*{\ch}[1]{\ensuremath{\mathrm{ch}_{#1}}}  
\NewDocumentCommand{\gchar}{ m o g }{\ensuremath{\chi_{#1}\IfNoValueF{#2}{^{#2}}\IfNoValueF{#3}{(#3)}}} 
\NewDocumentCommand{\grep}{ m o g }{\ensuremath{\rho_{#1}\IfNoValueF{#2}{^{#2}}\IfNoValueF{#3}{(#3)}}} 

\newcommand*{\CP}[0]{CP\xspace}

\newcommand*{\R}[0]{\ensuremath{R}\xspace}  

\newcommand*{\Z}[1]{\ensuremath{\mathbbm{Z}_{#1}}}  
\newcommand*{\inv}[0]{\ensuremath{^{-1}}}

\newcommand*{\conj}[0]{\ensuremath{^*}}
\newcommand*{\umat}[0]{\ensuremath{\mathbbm{1}}}  
\makeatletter
\newcommand*{\imod}[1]{\allowbreak\mkern10mu({\operator@font mod}\,\,#1)}
\makeatother

\DeclareMathOperator{\e}{e}  
\DeclareMathOperator{\I}{i}  
\DeclareMathOperator{\tr}{tr}  
\DeclareMathOperator{\rank}{rank}  

 \let\Phi\varPhi \let\varPhi\oldPhi
 \let\Psi\varPsi \let\varPsi\oldPsi
 \let\Lambda\varLambda \let\varLambda\oldLambda
\let\oldepsilon\epsilon \let\epsilon\varepsilon \let\varepsilon\oldepsilon

\newcommand{\authoraffill}[2]{%
{\sffamily\textbf{#1}
\\[3mm]
\textup{\small #2}}
\\[5mm]
}
\newcommand*{\email}[1]{\mdseries Email: \texttt{#1}}

\newcommand*{\ie}[0]{i.e.\ }
\newcommand*{\eg}[0]{e.g.\ }
\newcommand*{\cf}[0]{cf.\ }

\newcommand*{\Eqref}[1]{equation~\eqref{#1}}

\newcommand*{\af}[0]{\ensuremath{\mathrm{A}_4}\xspace}
\newcommand*{\afive}[0]{\ensuremath{\mathrm{A}_5}\xspace}
\newcommand*{\sfour}[0]{\ensuremath{\mathrm{S}_4}\xspace}
\newcommand*{\tp}[0]{\ensuremath{\mathrm{T}'}\xspace}
\newcommand*{\dts}[0]{\ensuremath{\Delta{(27)}}\xspace}
\newcommand*{\dff}[0]{\ensuremath{\Delta{(54)}}\xspace}
\let\semidirect\rtimes

\begin{document}
\begin{titlepage}

\title{Breaking classical Lie groups to finite subgroups -- an automated approach}
\author{Maximilian Fallbacher}

\makeatletter
\hypersetup{pdftitle={\@title},pdfauthor={\@author}}
\makeatother

\begin{flushright}
  TUM-HEP 996/15\\
  FLAVOUR(267104)-ERC-103
\end{flushright}

\vspace*{1.0cm}

\begin{center}

\begin{spacing}{1.1}
  \usekomafont{title}\huge 

  \makeatletter
  \@title
  \makeatother
\end{spacing}

\vspace{0.5cm}


\vspace{1cm}

\renewcommand*{\thefootnote}{\alph{footnote}}

\authoraffill{Maximilian Fallbacher\footnote{\email{m.fallbacher@tum.de}}}
{Physik-Department T30, Technische Universität München,\\
James-Franck-Straße 1, 85748 Garching, Germany}

\end{center}

\vspace{1cm}

\begin{abstract}
  The decomposition of representations of compact classical Lie groups into representations of finite subgroups is discussed.
  A \textsc{Mathematica} package is presented that can be used to compute these branching rules using the Weyl character formula.
  For some low order finite groups including \af and \dts general analytical formulas are presented for the branching rules of arbitrary representations of their smallest Lie super-groups.
\end{abstract}

\setcounter{footnote}{0}

\end{titlepage}

\hrule
\tocloftpagestyle{scrheadings}
\tableofcontents
\vspace{\baselineskip}
\hrule
\bigskip

\section{Introduction}

The Standard Model of particle physics (SM) provides a highly accurate description of Nature.
However, there are still many questions to be answered.
Amongst others, the so-called flavour puzzle remains to be solved, \ie a satisfactory explanation of the family structure with its masses and mixing is still lacking.
One possible avenue towards a solution is provided by settings with non-abelian discrete flavour or horizontal symmetries, which, somewhat complementary to grand unified symmetries, relate the different families of the SM.
Many models using such symmetries have been built, especially for the lepton sector, where until some time ago the so-called tri-bi-maximal mixing pattern seemed to be in good agreement with observation (for reviews of such models see, for example, \cite{Altarelli:2010gt,Ishimori:2010au,Grimus:2011fk,Altarelli:2012ss,King:2013eh,Feruglio:2015jfa}).

Even if one is able to find an explanation for the flavour structure of the SM in terms of finite non-abelian symmetries, this is, of course, only the first step.
One would also like to explain the origin of these symmetries.
This problem is amplified because global symmetries are believed to be broken by gravitational effects \cite{Abbott:1989jw,Banks:2010zn}.
One possible consistent origin of these symmetries is provided by string theory \cite{Kobayashi:2006wq,Abe:2009vi,Nilles:2012cy,BerasaluceGonzalez:2012vb}.
However, although in principle highly predictive, string theory has not yet been entirely successful in obtaining unambiguous predictions that can be compared with present-day experimental data.
Another, more bottom-up possibility to obtain discrete symmetries that are protected from violation by gravitational effects is by breaking a (non-anomalous) continuous gauge group like \SU{N} \cite{Krauss:1988zc}.
Symmetries of this kind are known as discrete gauge symmetries.

The aim of this work is to aid the construction of models where a finite symmetry emerges from the spontaneous breaking of a continuous symmetry, \ie from breaking Lie groups to finite subgroups.
This is only possible if there is an irreducible representation of the Lie group that contains a trivial singlet of the subgroup.
A vacuum expectation value of this component may then break the Lie group to the desired subgroup.
Therefore, a procedure is outlined and implemented in the \textsc{Mathematica} \cite{Mathematica10} package \texttt{DecomposeLGReps} allowing to decompose the Lie group's representations into irreducible representations of the subgroup.

The decomposition of a representation of a finite group into representations of a subgroup proceeds via the scalar product of characters.
The same technique can, in principle, be used for Lie groups as long as their dimension is finite.
However, whereas character tables of finite groups contain all information needed to go through this procedure, it is clearly impossible to compile all the necessary information for Lie groups; the table had to be infinitely large.

Hence, previous studies used different methods to compute the branching rules.
A first possibility is using the fact that each Lie group representation can be obtained from the tensor product of fundamental representations%
\footnote{So-called spinor representation of \SO{N} are an exception, see \autoref{sec:criteria} below.}
as in \cite{Luhn:2011ip}.%
\footnote{The title of the present work is an allusion to the title of this reference: \citetitle{Luhn:2011ip} by \citeauthor{Luhn:2011ip}.}
A second option is working with explicit realisations of the Lie group representations \cite{Adulpravitchai:2009kd}.
Both approaches cannot be easily generalised to larger Lie group representations or larger rank Lie groups.
In another approach that also highlights the connection between spontaneous and explicit symmetry breaking, \textcite{Merle:2011vy} used an algorithm based on group invariants and provided a \textsc{Mathematica} package implementing the algorithm for \SU{3}.
Again this is not easily generalised, and the method relies on somewhat advanced notions of invariant theory.
Similar considerations also lead to the so-called generating functions for subgroup scalars compiled in \cite{King:1982sj}, which, however, mainly focuses on Lie subgroups of Lie groups.
To overcome these limitations, the present work uses the standard technique of the character scalar product and computes the characters on the fly with \textsc{Mathematica} using what is called the Weyl character formula \cite{Weyl1953} (for more modern treatments see, for example, \cite{Fuchs2003,Fulton2004}).
This, in principle, allows the computation of branching rules for all compact Lie groups and arbitrary finite subgroups thereof.
The corresponding \textsc{Mathematica} package \texttt{DecomposeLGReps} implementing the formulas for the classical Lie groups \UU{N}, \SU{N}, \SO{N} and \USp{N} can be found \href{http://einrichtungen.ph.tum.de/T30e/codes/DecomposeLGReps}{online}.%
\footnote{\url{http://einrichtungen.ph.tum.de/T30e/codes/DecomposeLGReps}}

Of course, there remain some general issues with this type of model building.
For example, the VEV of the singlet component of the Lie group representation under consideration may be left invariant by a larger number of transformations than the desired subgroup, \ie the subgroup might not be the maximal invariant subgroup of the VEV.
Unfortunately, there is no general theory that exposes whether this is the case or not; hence, this question has to be settled in each case individually, \eg by examining the subgroup tree \cite{Merle:2011vy} or by constructing the actual representation matrices \cite{Luhn:2011ip,Adulpravitchai:2009kd}.
Furthermore, it is, in general, difficult to write down a potential giving rise to the desired VEV dynamically.
These caveats notwithstanding, knowledge of candidate representations for the desired breaking is of great help in model building.
Moreover, as will be seen below, in some cases one can discern patterns in the branching rules that allow to make general statement about models embodying this breaking.

This work is structured as follows.
First, criteria for a finite group to be a subgroup of a compact classical Lie group are compiled in \autoref{sec:criteria}.
In \autoref{sec:characterScalarProduct} the scalar product of characters and its application for the computation of branching rules is reviewed.
\autoref{sec:lieCharacters} explains the technical details of the computation, which uses the connection between Lie algebra and Lie group characters and the Weyl character formula.
The \textsc{Mathematica} package \texttt{DecomposeLGReps} implementing this procedure is briefly presented in \autoref{sec:package}.
In \autoref{sec:examplesForSmallGroups}, some examples are given for the applicability of the package, and some general results for various small finite groups are derived.

\section{Subgroups of compact classical Lie groups}
\label{sec:criteria}

The purpose of this section is to state criteria for a finite group $H$ to be a subgroup of any of the compact classical Lie groups \UU{N}, \SU{N}, \SO{N} and \USp{N}.
In fact, the simplest case is the one of the unitary group \UU{N} because any finite-dimensional representation of a finite group is equivalent to a unitary representation \cite{Fuchs2003}.

The criterion used here is that $H$ is a subgroup of \UU{N} if and only if it has a faithful representation of dimension $N$. 

To see this, let $H$ have a faithful unitary representation \rep{R} of dimension $N$,
\begin{equation}
  \rep{R}:~H \rightarrow \left\{ N \times N \text{ unitary matrices} \right\}\,,
\end{equation}
and let \rep{N} be the fundamental representation of \UU{N},
\begin{equation}
  \rep{N}:~\UU{N} \rightarrow \left\{ N \times N \text{ unitary matrices}\right\}\,,
\end{equation}
which is a bijection.
Then the map
\begin{equation}
  \rep{N}\inv\,\circ\,\rep{R}:~H \hookrightarrow \UU{N}
\end{equation}
is an injective group homomorphism that embeds $H$ into \UU{N} as a subgroup.

Now let $H$ be a subgroup of \UU{N}.
Then there exists an embedding $i$ of $H$ into \UU{N},
\begin{equation}
  i:~H \hookrightarrow \UU{N}\,,
\end{equation}
where $i$ is an injective group homomorphism.
Using this map one can define a faithful representation \rep[']{R} of $H$ by
\begin{equation}
  \rep[']{R} \coloneqq \rep{N}\,\circ\,i\,,
\end{equation}
which has dimension $N$.
This concludes the proof of the subgroup criterion for \UU{N}.

The same arguments go through for the other compact classical Lie groups if, additionally, $\det{\grep{\rep{R}}(g)}=1$ for \SU{N}, $\det{\grep{\rep{R}}(g)}=1$ and for some choice of basis $\grep{\rep{R}}{g} \in \mathbb{R}^{N \times N}$ for \SO{N} and $\grep{\rep{R}}{g} \in \Sp{N,\mathbb{C}}$ for \USp{N}.

For works on subgroups of the probably most relevant Lie groups for model building, \SU{3} and \SO{3}, see \cite{Grimus:2011fk,Ludl:2011gn,Ludl:2009ft,Fairbairn:1964sga}, and for a general overview of popular groups for model building, see \cite{Ishimori:2010au}.

One more important remark concerns the notation for Lie group representations used here.
An irreducible representation of a Lie group is labelled by the Dynkin labels $\Lambda=(\Lambda^1,\,\Lambda^2,\dots,\Lambda^r)$ of the highest weight of its associated Lie algebra representation.
This correspondence between Lie group and Lie algebra representations is only one-to-one for simply connected Lie groups, \eg for \SU{N} and \USp{N} (\cf \cite{Fuchs2003}).
However, \SO{N} is not simply connected and has as universal covering group $\mathrm{Spin}(N)$, \eg the universal covering group of \SO{3} is $\mathrm{Spin}(3)$ which is isomorphic to \SU{2}.
Thus, for the present approach one has to distinguish the groups \SO{N} and $\mathrm{Spin}(N)$ carefully in contrast to common practice in physics.
In fact, the $N$-dimensional fundamental representation of \SO{N} is not a faithful representation for $\mathrm{Spin}(N)$.
Therefore, the procedure described above really embeds the finite group into \SO{N} and not into $\mathrm{Spin}(N)$.
Branching rules will, hence, only be computed for non-spinorial representations, \ie for representations with Dynkin label $\Lambda^N$ an even integer for \SO{2N+1} and $\Lambda^{N-1}+\Lambda^N$ an even integer for \SO{2N}.

\section{Group characters}
\label{sec:characterScalarProduct}

This section reviews the notion of group characters.
More information can be found in many books on group theory, \eg \cite{Ramond2010}.

Let $G$ be a compact group and consider its finite-dimensional representations over the complex numbers.
Given such a representation \rep{R} with matrix realisation $\grep{\rep{R}}{g}$ one can define its character as
\begin{align}
  \gchar{\rep{R}}{g} & \coloneqq \tr{(\grep{\rep{R}}{g})}\,, \qquad \forall \, g \in G\,,
\end{align}
which is a map from the group to the complex numbers.
Since the trace is invariant under similarity transformations, characters are independent of the chosen basis.
Moreover, they are class functions, \ie they are constant on conjugacy classes.
Characters fulfil the relations
\begin{subequations}
\begin{align}
  \gchar{\rep{R}}{g\inv} & = \gchar{\rep{R}}{g}\conj\,, \qquad \forall \, g \in G\,,\\
  \gchar{\rep{R} \oplus \rep[']{R}}{g} & = \gchar{\rep{R}}{g} + \gchar{\rep[']{R}}{g}\,, \qquad \forall \, g \in G\,,\\
  \gchar{\rep{R} \otimes \rep[']{R}}{g} & = \gchar{\rep{R}}{g} \cdot \gchar{\rep[']{R}}{g}\,, \qquad \forall \, g \in G\,.
\end{align}
\end{subequations}
The characters of irreducible representations are in one-to-one correspondence with the representations themselves.

One can define a scalar product on characters by%
\footnote{For compact Lie groups the sum has, in principle, to be replaced by a suitably normalised integral.
Since we only need the character scalar product of finite subgroups, the present discussion is sufficient for our purposes.}
\begin{align}\label{eq:characterProduct}
  (\gchar{\rep{R}},\gchar{\rep[']{R}}) & \coloneqq \frac{1}{|G|}\, \sum_{g\in G}\,\gchar{\rep{R}}{g\inv} \, \gchar{\rep[']{R}}{g}  = \frac{1}{|G|}\, \sum_{g\in G}\,\gchar{\rep{R}}{g}\conj \, \gchar{\rep[']{R}}{g}\,,
\end{align}
where $|G|$ is the number of elements of $G$.
Given characters of two irreducible representations, their scalar product is $1$ if the representations are equivalent and $0$ if they are different, \ie the characters of irreducible representations are orthonormal.
Given a reducible representation \rep[_\mathrm{red}]{R}, the number of times the irreducible representation \rep[_\mathrm{irr}]{R} is contained in \rep[_\mathrm{red}]{R} is equal to $(\gchar{\rep[_\mathrm{irr}]{R}},\gchar{\rep[_\mathrm{red}]{R}})$.

This leads to the following algorithm for the computation of branching rules.
Given an irreducible representation \rep{R} of $G$, restrict it to a subgroup $H \subset G$.
Then the character scalar product of this representation with an irreducible representation \rep[_i]{r} of $H$ yields the multiplicity $\mu_i$ of \rep[_i]{r} in \rep{R},
\begin{align}\label{eq:characterScalarProductBranching}
  \mu_i = (\gchar{\rep[_i]{r}}, \left.\gchar{\rep{R}}\right|_H) =  \frac{1}{|H|}\, \sum_{h\in H}\,\gchar{\rep[_i]{r}}{h}\conj \, \gchar{\rep{R}}{h}\,.
\end{align}
This will now be applied to finite subgroups of Lie groups.
Note that, since the sum only runs over the finite number of elements of the subgroup, only a finite number of Lie group characters has to be computed.

\section{Lie group characters}
\label{sec:lieCharacters}

This section explains the details of the computation of Lie group characters using the Weyl character formula.
It is structured as follows.
First, the connection of Lie group and Lie algebra characters is reviewed, and the Weyl character formula for the computation of Lie algebra characters in its modern formulation is introduced.
After clarifying some notational issues, the Weyl character formulas for the classical Lie groups are presented in two formulations due to Weyl, which are for the present purposes more useful than the general formula mentioned before.
The reader only interested in the application of the \textsc{Mathematica} package or the final results for small groups may skip ahead to the respective sections.

\subsection*{Lie group and Lie algebra characters}

The definition of characters shown above is not limited to finite groups.
In fact, the character \gchar{\Lambda} of a finite-dimensional highest-weight representation of some finite-dimensional Lie group $L$ is defined in the same way, namely
\begin{align}
  \gchar{\Lambda}(g) \coloneqq \tr(\grep\Lambda(g))\,, \quad \forall g \in L\,,
\end{align}
where $\grep\Lambda$ is a matrix realisation of the representation with highest weight $\Lambda$ \cite{Fuchs2003}.
The characters are again class-functions, \ie constant on conjugacy classes.

It is important to note that for semi-simple Lie groups any group element $g$ is conjugate to an element $g_\circ$ of a maximal torus, \ie of a subgroup with the Cartan sub-algebra of $L$ as Lie algebra.
In other words: each group element can be diagonalised by an inner automorphism \cite{Fuchs2003}.
This leads to a relation between so-called Lie algebra characters%
\footnote{Lie algebra characters \ch{\Lambda} are defined as \cite{Fuchs2003}
\begin{align}
  \ch{\Lambda}(\lambda) \coloneqq \sum_{\lambda'} \operatorname{mult}_\Lambda(\lambda')\,\e^{(\lambda,\,\lambda')}\,,
\end{align}
where the sum runs over all weights $\lambda'$ of the representation defined by $\Lambda$ and $\operatorname{mult}_\Lambda(\lambda')$ is the multiplicity of $\lambda'$ in the weight diagram.
Moreover, the parenthesis denote the scalar product on weight space.}
\ch{\Lambda} and the desired Lie group characters \gchar{\Lambda}. 
In fact, for each element $g_\circ$ of a maximal torus of a compact Lie group $L$, one can find an element $h$ of the Cartan sub-algebra of the Lie algebra of $L$ such that $\exp{h}=g_\circ$.
The Lie algebra character $\ch{\Lambda}(h)$ of this element equals the Lie group character of $g$ \cite{Fuchs2003},
\begin{align}
  \ch{\Lambda}(h) = \gchar{\Lambda}{\exp{h}} = \gchar{\Lambda}{g_\circ} = \gchar{\Lambda}{g}\,.
\end{align}

It is, hence, possible to compute all Lie group characters using this equivalence with Lie algebra characters given a formula for the latter.
In fact, there is a closed formula for Lie algebra characters called Weyl character formula, which in its modern form is given by \cite{Fuchs2003}
\begin{align}\label{eq:WeylCharacterFormula}
  \ch{\Lambda}(h) = \ch{\Lambda}\left(\lambda_h\right) = \frac{\sum_{w\in W} \operatorname{sign}(w)\,\e^{(w(\Lambda+\rho),\,\lambda_h)}}{\sum_{w\in W} \operatorname{sign}(w)\,\e^{(w(\rho),\,\lambda_h)}}\,.
\end{align}
This formula requires some explanation.
Note that the element $h$ of the Cartan sub-algebra for which the character is computed is specified by its weight $\lambda_h$, which can be obtained by projecting $h$ onto the Cartan generators using the Killing form.
Further, the sums run over all elements of the Weyl group $W$.
This is the group generated by all reflections in weight space at planes orthogonal to the simple roots, \ie it is generated by all so-called Householder transformations corresponding to the simple roots.
The sign of an element $w$ of the Weyl group is defined as $\operatorname{sign}(w) \coloneqq (-1)^{\operatorname{length}(w)}$, where the length of an element is the (unique) minimal number of reflections defined by simple roots that is needed to generate the reflection $w$.
The structures of the Weyl groups of the classical Lie algebras are shown in \autoref{tab:WeylGroup}.
\begin{table}
\centering
\begin{tabular}{lcc}
\toprule
  \lalg{g} & $W$ & $|W|$\\
\midrule
  $A_r$ & $\mathrm{S}_{r+1}$ & $(r+1)!$\\
  $B_r$ & $\Z{2}^r\semidirect \mathrm{S}_r $ & $2^r\,r!$\\
  $C_r$ & $\Z{2}^r\semidirect \mathrm{S}_r $ & $2^r\,r!$\\
  $D_r$ & $\Z{2}^{r-1}\semidirect \mathrm{S}_r $ & $2^{r-1}\,r!$\\
\bottomrule
\end{tabular}
\caption{This table is a partial reproduction of \cite[(10.29)]{Fuchs2003} and shows the group structures and orders of the Weyl groups of the classical Lie algebras.}
\label{tab:WeylGroup}
\end{table}
Moreover, $(\cdot\,,\,\cdot)$ is the scalar product on weight space and $\rho$ denotes the Weyl vector of the Lie algebra, which is given by half the sum of the positive roots,
\begin{align}
  \rho \coloneqq \frac{1}{2}\, \sum_{\alpha>0} \alpha\,.
\end{align}
It has Dynkin labels $\rho^i=1$ for all $i=1,\dots,\rank{L}$.

\subsection*{Notation for the Weyl character formulas in terms of eigenvalues}

The Weyl character formula will finally be applied to Lie group elements that are also elements of the finite subgroup in order to compute the branching rules using the character scalar product \eqref{eq:characterScalarProductBranching}.
In the case of embeddings as described in \autoref{sec:criteria}, these Lie group elements $g$ are not specified by weights but by an explicit representation matrix \grep{\rep{N}}{g} in the fundamental representation.
Instead of first translating this information into the language of weights, one can also compute the characters directly in terms of the eigenvalues $\epsilon_i(g)$ of these representation matrices \cite{Weyl1953,Proctor1994,Fulton2004}. 
In this case, the irreducible Lie group representation whose character is to be computed is conventionally not labelled by Dynkin labels but by its so-called partition, a notation related to Young tableaux.
For \SU{N}, \ie Lie algebra $A_{r=N-1}$, the relation of Dynkin labels $\Lambda=(\Lambda^1,\,\Lambda^2,\dots,\Lambda^{N-1})$ to partitions can be understood from the fact that $\Lambda^i$ is the number of columns with  $i$ boxes in the Young tableau corresponding to $\Lambda$.
The partition is then the list of row lengths $f_i$ of the Young tableau, which can be computed from the Dynkin labels by
\begin{align}\label{eq:flabels}
  f_i \coloneqq \sum_{k=i}^{N-1}\,\Lambda^k\,, \quad i=1,\dots,N-1\,,
\end{align}
and which results naturally in the ordering $f_i\geq f_{i+1}$ \cite{Proctor1994,Fulton2004}.

Partitions are also the conventional way to label irreducible representations of \UU{N} \cite{Weyl1953}. 
However, in this case an additional integer $f_N$ has to be specified.
Furthermore, all $f_{i\leq N-1}$ as computed with the formula above have to be increased by this $f_N$.
Restricting representations from \UU{N} to its \SU{N} subgroup, all representations differing only in this global shift are identical and $f_N$ can be set to zero without loss of generality.

For representations $\Lambda$ of the symplectic group \USp{N}, \ie Lie algebra $C_{r=N}$, the $N$ labels $f_i$ of the corresponding Young tableau are again obtained by \Eqref{eq:flabels} with the sum extending up to $N$ this time \cite{Fulton2004}.

The definition of Young tableaux and partitions for orthogonal groups is more complicated and not unique; thus, one has to be careful when comparing different approaches.
In addition to that, one has to distinguish \SO{2N}, \ie Lie algebra $D_{r=N}$, and \SO{2N+1}, \ie Lie algebra $B_{r=N}$.
We adopt the conventions of \cite{Proctor1994,Fulton2004}.%
\footnote{The conventions by \textcite{Weyl1953} differ only slightly from the other two.
He uses the absolute value of $f_N$ for \SO{2N} and adds a prime to distinguish between representations with positive and negative $f_N$.}
Hence, the partition corresponding to $\Lambda$ is obtained from
\begin{subequations}
\allowdisplaybreaks
\begin{align}
  f_i & \coloneqq \sum_{k=i}^{N-1}\,\Lambda^k + \frac{\Lambda^N}{2}\,, \quad i=1,\dots,N\,, \quad \text{for \SO{2N+1} and}\\
  f_i & \coloneqq \sum_{k=i}^{N-2}\,\Lambda^k + \frac{\Lambda^{N-1} + \Lambda^{N}}{2}\,, \quad i=1,\dots,N-1\,, \quad f_N\coloneqq \frac{\Lambda^N - \Lambda^{N-1}}{2}\,, \quad \text{for \SO{2N},}
\end{align}
\end{subequations}
where the sums are set to zero if $i$ is larger than their upper limit.
All $f_i$ are integers for non-spinorial representations but half-integer for spinor representations.
This does not pose any problem for the present approach since only subgroups of \SO{N} are considered, and, therefore, only non-spinorial representations are allowed as input.

It will turn out to be convenient to furthermore introduce the abbreviation
\begin{align}\label{eq:defEll}
  \ell_i & \coloneqq f_i - i + N\,, \quad i=1,\dots,N\,,
\end{align}
setting $f_N\coloneqq 0$ for \SU{N}.

\subsection*{The Weyl character formulas in terms of eigenvalues}

After introducing this notation, the character formulas simply take the form of determinants.
Taking \SU{N} as an example, this can be seen starting from \eqref{eq:WeylCharacterFormula}.
The sum over the Weyl group of the signum of the Weyl group elements times an exponential resembles, the Weyl group of \SU{N} being $S_{N}$, the Leibniz formula for the determinant of a matrix.
After some algebraic manipulations one can indeed write both numerator and denominator as determinants.
Further, the weights corresponding to a group element $g$ can be related to the eigenvalues $\epsilon_i(g)$ of its representation matrix in the fundamental representation.
The final result for the Weyl character formula for \SU{N} in terms of these eigenvalues $\epsilon_i(g)$ using the $\ell_i$ introduced in \eqref{eq:defEll} is then \cite{Weyl1953}
\begin{align}
  \gchar{\Lambda}{g} = \frac{\det{\left[ \epsilon_i^{\ell_j}(g) \right]_{ij}}}{\det{\left[ \epsilon_i^{N-j}(g) \right]_{ij}}}\,.
\end{align}
Here, $\det[A]_{ij}$ is the determinant of the $n\times n$ matrix $A$ with entries labelled by $1\leq i,j\leq n$.
This expression is also called a Schur polynomial \cite{Fulton2004}.
In fact, the denominator can be simplified because it is just a Vandermonde determinant, yielding
\begin{align}
  \gchar{\Lambda}(g) = \frac{\det{\left[ \epsilon_i^{\ell_j}(g) \right]_{ij}}}{\prod_{i<j} \left(\epsilon_i(g) - \epsilon_j(g)\right) }\,.
\end{align}
This formula holds also for \UU{N} \cite{Weyl1953}.

The other compact classical Lie groups can be treated similarly.
However, in all these cases only half of the eigenvalues are independent because they always come in complex conjugate pairs.%
\footnote{Matrices of \SO{2N+1} have an additional eigenvalue $+1$ which also has to be dropped from the list.}
Hence, for all groups besides the unitary groups, only one eigenvalue of each pair is to be used in the formulas below such that their number matches the rank of the Lie algebra.
The formulas for all compact classical Lie groups are then \cite{Weyl1953,Fulton2004}
\begin{subequations}\label{eq:WCFeigenvalues}
\allowdisplaybreaks
\begin{align}
  \gchar{\Lambda}(g) & = \frac{\det{\left[ \epsilon_i^{\ell_j}(g) \right]_{ij}}}{\prod_{i<j} \left(\epsilon_i(g) - \epsilon_j(g)\right) }\, && \text{for \SU{N},}\\
  \gchar{\Lambda}(g) & = \frac{\det{\left[ \epsilon_i^{\ell_j+1}(g) - \epsilon_i^{-\ell_j-1}(g)\right]_{ij}}}{\det{\left[ \epsilon_i^{N+1-j}(g) - \epsilon_i^{-N-1+j}(g) \right]_{ij}}}\, && \text{for \USp{N},} \label{eq:WCFeigenvaluesUSpn}\\
  \gchar{\Lambda}(g) & = \frac{\det{\left[ \epsilon_i^{\ell_j+1/2}(g) - \epsilon_i^{-\ell_j-1/2}(g)\right]_{ij}}}{\det{\left[ \epsilon_i^{N+1/2-j}(g) - \epsilon_i^{-N-1/2+j}(g) \right]_{ij}}}\, && \text{for \SO{2N+1},}\\
  \gchar{\Lambda}(g) & = \frac{\det{\left[ \epsilon_i^{\ell_j}(g) + \epsilon_i^{-\ell_j}(g)\right]_{ij}}+\det{\left[ \epsilon_i^{\ell_j}(g) - \epsilon_i^{-\ell_j}(g)\right]_{ij}}}{\det{\left[ \epsilon_i^{N-j}(g) + \epsilon_i^{-N+j}(g) \right]_{ij}}}\,&& \text{for \SO{2N}.} \label{eq:WCFeigenvaluesSO2np1}
\end{align}
\end{subequations}
These formulas are implemented in the \textsc{Mathematica} package \texttt{DecomposeLGReps}.

Unfortunately, there is a computational difficulty because all determinants are zero if any two eigenvalues coincide.
This can be most easily seen in the case of \SU{N}, where the Vandermonde determinant clearly vanishes for two identical eigenvalues.
Fortunately, this is just a removable discontinuity.
In the original formula \eqref{eq:WeylCharacterFormula} this can be ameliorated by adding a multiple of the Weyl vector $t\cdot\rho$ to the weight $\lambda$ and taking the limit $t \to 0$ after computing the determinant.
In \eqref{eq:WCFeigenvalues} the same can be achieved by the replacement $\epsilon_j \to \epsilon_j\,\e^{\I j t}$ and the limit $t\to 0$.

The formulas are computationally rather demanding because of the possibly large determinants.
Computation time should roughly grow as $(r+1)!$, where $r$ is the rank of the Lie group.
However, for the ranks of Lie groups usually used in model building this is not a major concern.

A big advantage of these formulas is that they are closed, \ie they do not involve any recursion in contrast to, for example, the Freudenthal formula \cite{Fuchs2003}.
Hence, they can be used to derive general properties for subgroups of classical Lie groups, see \autoref{sec:examplesForSmallGroups} below.

\subsection*{An alternative formulation of the Weyl character formulas}

If one only needs a result for fixed integer Dynkin labels, a second form of the character formulas can be advantageous.
This form circumvents the limit procedure, which, otherwise, considerably slows down the computation.
It can be derived using a correspondence between Schur polynomials and determinants of complete homogeneous symmetric polynomials $h_i$, which are defined by
\begin{align}
  \frac{1}{\prod_i{(1 - z\, x_i)}} \eqqcolon \sum_j h_j(x_i)\,z^j\,,
\end{align}
see \cite{Weyl1953,Fulton2004}.
In the present case the $h_i$ are to be evaluated at the eigenvalues of the representation matrix.
In fact, the quantities from which the characters can be computed are the coefficients $p_i$ of the generating function for one divided by the characteristic polynomial of this matrix \cite{Weyl1953,Fulton2004},
\begin{align}\label{eq:homogeneousPol}
  \frac{1}{\det{(\umat - z\, \grep{\Lambda}(g))}} = \frac{1}{\prod_i{(1 - z\, \epsilon_i(g))}}= \sum_j h_j(\epsilon_i(g))\,z^j \eqqcolon \sum_j p_j(g)\,z^j\,.
\end{align}
The final formulas for the characters of \UU{N}, \SU{N} and \USp{N} are given by \cite{Weyl1953}
\begin{subequations}\label{eq:WCFcharpol}
\begin{align}
  \gchar{\Lambda}(g) & = \det{\left[ p_{\ell_i - N + j}(g)\right]_{ij}}&& \text{for \UU{N} and \SU{N},}\\
  \gchar{\Lambda}(g) & = \frac{1}{2}\,\det{\left[ p_{\ell_i - N + j}(g) + p_{\ell_i - N - j + 2}(g) \right]_{ij}}&& \text{for \USp{N},} \label{eq:WCFcharpolUSpn}
\end{align}
Formulas for \SO{N} cannot be found in \cite{Weyl1953}, but for \OO{2N} and \OO{2N+1}
\begin{align}
  \gchar{\Lambda}(g) & = \det{\left[ p_{\ell_i - N + j}(g) - p_{\ell_i - N - j}(g) \right]_{ij}}\,.\notag
\end{align}
The irreducible representations of \SO{2N+1} and \OO{2N+1} coincide such that the character formula for \OO{2N+1} can also be used for \SO{2N+1}.
However, only irreducible representations of \SO{2N} whose last two Dynkin labels are equal are also irreducible representations of \OO{2N}, in which case the characters are again identical.
Irreducible representations of \SO{2N} with Dynkin labels $\Lambda^{N-1} \neq \Lambda^N$ are not representations of \OO{N}.
Instead, the direct sum of the two conjugate representations $(\Lambda^1,\dots,\,\Lambda^{N-1},\,\Lambda^N)$ and $(\Lambda^1,\dots,\,\Lambda^{N},\,\Lambda^{N-1})$ of \SO{2N} forms an irreducible representation of \OO{2N} \cite{Fulton2004}.
Using a determinant formula from \cite{Proctor1994} on \eqref{eq:WCFeigenvaluesSO2np1} and comparing with \eqref{eq:WCFeigenvaluesUSpn} one can derive the formula for the remaining representations of \SO{2N}.
It depends on the sign of $\Lambda^{N-1} - \Lambda^N$.
In summary, the results for \SO{N} are
\enlargethispage{2\baselineskip}
\begingroup
\allowdisplaybreaks
\begin{align}
  \gchar{\Lambda}(g) & = \det{\left[ p_{\ell_i - N + j}(g) - p_{\ell_i - N - j}(g) \right]_{ij}}&& \text{for \SO{2N+1},}\\
  \gchar{\Lambda}(g) & = \det{\left[ p_{\ell_i - N + j}(g) - p_{\ell_i - N - j}(g) \right]_{ij}} && \text{for \SO{2N} with } \Lambda^{N-1}=\Lambda^N\,,\\
  \gchar{\Lambda}(g) & = \frac{1}{2}\,\det{\left[ p_{\ell_i - N + j}(g) - p_{\ell_i - N - j}(g) \right]_{ij}} + \notag\\
    & \qquad + \mathrlap{\frac{\operatorname{sign}(\ell_N)}{4}\, \prod_k(\epsilon_k(g) - \epsilon_k(g)\inv)\, \det{\left[ p_{\ell_i - N + j - 1}(g) + p_{\ell_i - N - j + 1}(g) \right]_{ij}}}\notag\\
    & && \text{for \SO{2N} with, } \Lambda^{N-1}\neq \Lambda^N\,.
\end{align}
\endgroup
\end{subequations}
The Weyl character formulas thus obtained can be implemented on a computer with the help of a computer algebra system like \textsc{Mathematica} which provides routines for the series computation \eqref{eq:homogeneousPol} needed to determine the $p_i$.
This has been done in the package \texttt{DecomposeLGReps} presented in the following section.

\section{The package}
\label{sec:package}

The \textsc{Mathematica} package \texttt{DecomposeLGReps} can be found on the webpage
\begin{center}
  \url{http://einrichtungen.ph.tum.de/T30e/codes/DecomposeLGReps}
\end{center}
It contains implementations of the Weyl character formulas \eqref{eq:WCFeigenvalues} as well as of the alternative form \eqref{eq:WCFcharpol}.
For a detailed explanation of the functions and their options, the reader is referred to the package documentation shipped with the package.
Here only the basic usage is briefly explained.

After loading the package with
\begin{verbatim}
  Needs["DecomposeLGReps`"];
\end{verbatim}
one has to specify the finite group that is to be embedded into a Lie group.
This is done by providing a list containing one list for each irreducible representation of the finite group with the representation matrices of all group elements.
Schematically this looks like
\begin{verbatim}
  group = { { list of representation matrices of representation 1 },
            { list of representation matrices of representation 2 },
            ...
            { list of representation matrices of representation n } };
\end{verbatim}
This list can, for example, be computed with the GAP interface package \texttt{Discrete} \cite{Holthausen:2011vd}.
Alternatively, it is also possible just to specify representation matrices for one representative of each conjugacy class, see the package documentation for more information.

After this preparation, the finite group can be embedded into a Lie group using \texttt{embedinLG},
\begin{verbatim}
  embed = embedinLG[group, 12, "A"];
\end{verbatim}
where the first argument is the list prepared before, the second argument specifies the representation that is used for the embedding following \autoref{sec:criteria}, and the last argument specifies the Lie group type.%
\footnote{Possible types are \texttt{"A"} for \SU{N} and \UU{N}, \texttt{"B"} for \SO{2N+1}, \texttt{"C"} for \USp{N} and \texttt{"D"} for \SO{2N}.}
If a reducible representation is to be embedded, a list of its irreducible constituents can be provided instead of a single integer as second argument.
Hence, in the example the group is embedded into $\SU{N}\sim A_{N-1} \sim$ \texttt{"A"} using the 12th representation in the list \texttt{group}, where $N$ is automatically chosen as the dimension of representation number $12$.
The representation chosen should, of course, be faithful; otherwise, the embedded group is not the desired one but a subgroup of it.
An error is displayed if this is detected.

The last step is to compute the decomposition of a representation of the Lie group specified by the Dynkin labels of its highest weight.
This is done by the function \texttt{decomposeLGRep} in the following way:
\begin{verbatim}
  decomposeLGRep[{a1, a2,..., aN}, embed]
\end{verbatim}
The first argument is a list with the Dynkin labels and the second argument is the output of \texttt{embedinLG}.
The Lie group type is also taken from there in order to avoid a mismatch between the Lie group of the embedding and the Lie group for which the branching rule is to be computed.
The output of \texttt{decomposeLGRep} is a list containing the multiplicities of representations of the finite group in the decomposition of the Lie group representation with the Dynkin labels $(\texttt{a1},\,\texttt{a2},\dots,\texttt{aN})$.
The order of the multiplicities in the output is identical to the one of representations $1$ to $n$ specified earlier in the variable \texttt{group} .

As an example, let \texttt{a4Matrices} contain the representation matrices of the tetrahedral group \af in the form shown above and in the order $(\rep{1},\,\rep[']{1},\,\rep['']{1},\,\rep{3})$ where the notation of \cite{Ludl:2009ft} is used.
The tetrahedral group can be embedded into \SU{3} using the faithful triplet representation \rep{3}.%
\footnote{In fact, it is a subgroup of \SO{3}, see the following section.}
This is done by the command
\begin{verbatim}
  embedA4 = embedinLG[a4Matrices, 4, "A"];
\end{verbatim}
To avoid confusion with the group name \af, let us remark that the $4$ stands for the fourth representation in the list \texttt{a4Matrices}, which is assumed to be ordered as $(\rep{1},\,\rep[']{1},\,\rep['']{1},\,\rep{3})$, and \texttt{"A"} for the Lie algebra of \SU{N}.
The decomposition of the fundamental representation of \SU{3} can then be computed by
\begin{verbatim}
  decomposeLGRep[{1, 0}, embedA4]
\end{verbatim}
which yields
\begin{verbatim}
  {0, 0, 0, 1}
\end{verbatim}
\ie the fundamental of \SU{3} contains once the \rep{3} of \af and no other representation.
This just shows that the embedding worked out correctly.
One can now compute more branching rules, \eg
\begin{verbatim}
  decomposeLGRep[{2, 0}, embedA4]  -> {1, 1, 1, 1}
  decomposeLGRep[{1, 1}, embedA4]  -> {0, 1, 1, 2}
  decomposeLGRep[{23, 15}, embedA4] -> {640, 640, 640, 1920}
\end{verbatim}

For more examples and explanations of all options, see the package manual included in the download.

Note that the package was checked for correctness against results for branching rules from the literature.
Indeed, all branching rules presented by \textcite{Luhn:2008sa,Luhn:2011ip} were reproduced successfully.
For the decompositions $\SO{3} \to \af$, $\SO{3} \to \sfour$ and $\SU{3} \to \dts$, this consistency check can easily be repeated by specialising the general formulas shown in the following section to the representations of smallest dimension.

\section{Examples for small finite groups}
\label{sec:examplesForSmallGroups}

Using the \textsc{Mathematica} package \texttt{DecomposeLGReps} presented in the previous section, one can derive general results for branching rules to some well-known finite groups.
This can be done by applying the character formulas \eqref{eq:WCFeigenvalues}, which allow for generic non-negative integer inputs for the Dynkin labels of the representations which are to be decomposed.
In all cases not only the exact functions determining the branching are interesting.
In addition, the insight gained on the structure, \ie on which representations are contained within which (congruence) class \cite{Lemire1980} of representations of the continuous group, is very helpful for model building.
The examples chosen are \af, \tp, \sfour, \afive, \dts and \dff.
Further information on all these groups can be found in \cite{Ishimori:2010au} although the notation used here is partly different.
References to the notations used are given for each case individually below.
In many cases, the results are actually independent of the specific naming convention, \eg in \af the results do not depend on which representation is called \rep[']{1} and which one \rep['']{1}.

The following abbreviations will be used for functions that occur several times:
\begin{subequations}
\begin{align}
  f(n,m) & \coloneqq (1 + n)\, (1 + 3 m + n)\, (2 + 3 m + 2 n)\,,\\
  p^+(n) & \coloneqq \cos{\left(\frac{n\,\pi}{3}\right)} + \frac{1}{\sqrt{3}} \sin{\left(\frac{n\,\pi}{3}\right)} = \begin{cases} 1, & n\equiv 0,1 \imod{6},\\ 0, & n \equiv 2,5 \imod{6},\\ -1, & n \equiv 3,4 \imod{6}, \end{cases}\\
  p^-(n) & \coloneqq \cos{\left(\frac{n\,\pi}{3}\right)} - \frac{1}{\sqrt{3}} \sin{\left(\frac{n\,\pi}{3}\right)} = \begin{cases} 1, & n\equiv 0,5 \imod{6},\\ 0, & n \equiv 1,4 \imod{6},\\ -1, & n \equiv 2,3 \imod{6}, \end{cases}\\
  q(n) & \coloneqq \cos{\left(\frac{4\,n\,\pi}{3}\right)} + \frac{1}{\sqrt{3}} \sin{\left(\frac{4\,n\,\pi}{3}\right)} = \begin{cases} 1, & n\equiv 0 \imod{3},\\ 0, & n \equiv 2 \imod{3},\\ -1, & n \equiv 1 \imod{3}. \end{cases}
\end{align}
\end{subequations}

\subsection{\texorpdfstring{\af}{A4}}

The tetrahedral group \af is very popular in model building because it can lead to the so-called tri-bi-maximal mixing structure for the neutrino mixing matrix \cite{Ma:2001dn,Altarelli:2005yp}.
It is a subgroup of \SO{3}; the embedding proceeds via the only three-dimensional representation \rep{3}.
The other representations are named as in \cite{Ludl:2009ft}.

The decomposition formulas are most easily displayed if the \SO{3} representations are split into five classes with Dynkin labels taking the forms $(12\,n+2m)$ for $m=0,\dots,5$.%
\footnote{Note that, since \af is a subgroup of \SO{3} not \SU{2}, only non-spinorial, \ie genuine, representations of \SO{3} are considered, see the discussion at the end of \autoref{sec:criteria}.}
The resulting multiplicities are displayed in \autoref{tab:A4}.
\begin{table}
\centering
\begin{tabular}{>{$}l<{$}@{~~$\to$~~}>{$}c<{$}>{$}c<{$}>{$}c<{$}>{$}c<{$}}
  \toprule
  \Lambda & \rep{1} & \rep[']{1} & \rep['']{1} & \rep{3}\\
  \midrule
  (12\,n) & n+1 & n & n & 3 n \\
  (12\,n+2) & n & n & n & 3 n+1 \\
  (12\,n+4) & n & n+1 & n+1 & 3 n+1 \\
  (12\,n+6) & n+1 & n & n & 3 n+2 \\
  (12\,n+8) & n+1 & n+1 & n+1 & 3 n+2 \\
  (12\,n+10) & n & n+1 & n+1 & 3 (n+1)\\
  \bottomrule
\end{tabular}
\caption{Branching rules for the embedding $\af \hookrightarrow \SO{3}$ using the triplet representation of \af.
\SO{3} representations are denoted by the Dynkin labels $\Lambda$ of their highest weights.
Only proper \SO{3} representations are considered, \ie $\Lambda$ is even, see \autoref{sec:criteria}.
For the conventions used, see \cite{Ludl:2009ft}.}
\label{tab:A4}
\end{table}
Setting $n$ to zero one obtains the branching rules for \SO{3} representations up to dimension $11$.
They are identical to the decomposition rules derived by \textcite{Luhn:2008sa}.

The smallest \SO{3} representation containing a trivial \af singlet is the representation with Dynkin label $(6)$, which using its dimension can also be denoted \rep{7}.

\subsection{\texorpdfstring{\tp}{T'}}

The double covering group \tp of \af is not a subgroup of \SO{3} but can be embedded into \SU{2} using the representation \rep[_0]{2}.
It can also lead to tri-bi-maximal neutrino mixing \cite{Chen:2007afa}.
The conventions are taken from \cite[Appendix~A.1]{Chen:2014tpa}.

Splitting the \SU{2} representations into the two classes of vector $(2\,n)$ and spinor $(2\,n+1)$ representations, the decomposition yields
\begin{subequations}
\allowdisplaybreaks
\begin{align}
  \begin{split}
   (2\,n) & \to \frac{1}{12}\, \left[2 n+(-1)^n\, \left(8\, p^-(n) +9\right)+1\right] \times \rep[_0]{1}\\
    & \qquad {}\oplus \frac{1}{12}\, \left[2 n+(-1)^n\, \left(-4\, p^-(n) +9\right)+1\right] \times (\rep[_1]{1} \oplus \rep[_2]{1})\\
    & \qquad {}\oplus \frac{1}{4}\, \left(2 n+(-1)^{n+1}+1\right) \times \rep{3}\,,
  \end{split}\\
  \begin{split}
   (2\,n+1) & \to \frac{1}{3}\, \left(n+2\, (-1)^n\, p^+(n) +1\right) \times \rep[_0]{2}\\
    & \qquad {}\oplus \frac{1}{3}\, \left(n+(-1)^{1+n}\, p^+(n) +1\right) \times (\rep[_1]{2} \oplus \rep[_2]{2})\,.
  \end{split}
\end{align}
\end{subequations}
In fact, the decomposition for vector representations is exactly the same as the one for $\af \hookrightarrow \SO{3}$ shown in \autoref{tab:A4} with the change of notation $\rep{1} \to \rep[_0]{1}$, $\rep[']{1} \to \rep[_1]{1}$ and $\rep['']{1} \to \rep[_2]{1}$.
For spinor representations the formulas above can be recast as shown in \autoref{tab:tp}.
\begin{table}
\centering
\begin{tabular}{>{$}l<{$}@{~~$\to$~~}>{$}c<{$}>{$}c<{$}>{$}c<{$}}
  \toprule
  \Lambda & \rep[_0]{2} & \rep[_1]{2} & \rep[_2]{2}\\
  \midrule
  (12\,n +1)    & 2 n+1 & 2 n & 2 n\\
  (12\,n+3)  & 2 n & 2 n+1 & 2 n+1\\
  (12\,n+5)  & 2 n+1 & 2 n+1 & 2 n+1\\
  (12\,n+7)  & 2 (n+1) & 2 n+1 & 2 n+1\\
  (12\,n+9)  & 2 n+1 & 2 (n+1) & 2 (n+1)\\
  (12\,n+11) & 2 (n+1) & 2 (n+1) & 2 (n+1)\\
  \bottomrule
\end{tabular}
\caption{Branching rules for the embedding $\tp \hookrightarrow \SU{2}$ using the doublet \rep[_0]{2} of \tp.
\SU{2} representations are denoted by the Dynkin labels $\Lambda$ of their highest weights.
Only spinor representations are considered because the branching rules for non-spinorial representations are the same as for \af shown in \autoref{tab:A4}.
For the \tp notation used, see \cite[Appendix~A.1]{Chen:2014tpa}.}
\label{tab:tp}
\end{table}
This shows that the doublet representations of \tp, which are not representations of \af, are ``spinor'' representations and can only be obtained from spinor representations of \SU{2}.
In particular, spinor representations of \SU{2} cannot be used to break \SU{2} to \tp because they do not contain trivial \tp singlets.

\subsection{\texorpdfstring{\sfour}{S4}}

The same classes as for \af can be used for \sfour, which is also a subgroup of \SO{3}.
It was used early on in flavour model building \cite{Yamanaka:1981pa} and is still popular because it, too, can lead to tri-bi-maximal mixing.
The embedding proceeds via representation \rep[']{3}.
The other three-dimensional representation \rep{3} would lead to an embedding into \OO{3} because not all its determinants are $+1$.
Again the notation from \cite{Ludl:2009ft} is used.

The results are shown in \autoref{tab:S4}.
\begin{table}
\centering
\tabulinesep=4pt
\begin{tabu}{>{$}l<{$}@{~~$\to$~~}>{$}c<{$}>{$}c<{$}>{$}c<{$}>{$}c<{$}>{$}c<{$}}
  \toprule
  \Lambda & \rep{1} & \rep[']{1} & \rep{2} & \rep{3} & \rep[']{3}\\
  \midrule
  (12\,n)   & \frac{2 n+(-1)^n+3}{4} & \frac{2 n+(-1)^{n+1}+1}{4} & n & \frac{6 n+(-1)^n-1}{4} & \frac{6 n+(-1)^{n+1}+1}{4} \\
  (12\,n+2) & \frac{2 n+(-1)^n-1}{4} & \frac{2 n+(-1)^{n+1}+1}{4} & n & \frac{6 n+(-1)^n+3}{4} & \frac{6 n+(-1)^{n+1}+1}{4} \\
  (12\,n+4) & \frac{2 n+(-1)^{n+1}+1}{4} & \frac{2 n+(-1)^n-1}{4} & n+1 & \frac{6 n+(-1)^{n+1}+1}{4} & \frac{6 n+(-1)^n+3}{4} \\
  (12\,n+6) & \frac{2 n+(-1)^{n+1}+1}{4} & \frac{2 n+(-1)^n+3}{4} & n & \frac{6 n+(-1)^{n+1}+5}{4} & \frac{6 n+(-1)^n+3}{4} \\
  (12\,n+8) & \frac{2 n+(-1)^n+3}{4} & \frac{2 n+(-1)^{n+1}+1}{4} & n+1 & \frac{6 n+(-1)^n+3}{4} & \frac{6 n+(-1)^{n+1}+5}{4} \\
  (12\,n+10)& \frac{2 n+(-1)^n-1}{4} & \frac{2 n+(-1)^{n+1}+1}{4} & n+1 & \frac{6 n+(-1)^n+7}{4} & \frac{6 n+(-1)^{n+1}+5}{4} \\
  \bottomrule
\end{tabu}
\caption{Branching rules for the embedding $\sfour \hookrightarrow \SO{3}$ using the triplet representation \rep[']{3} of \sfour.
\SO{3} representations are denoted by the Dynkin labels $\Lambda$ of their highest weights.
For the conventions used, see \cite{Ludl:2009ft}.}
\label{tab:S4}
\end{table}
The first trivial singlet occurs for the representation with Dynkin label $(8)$, which can also be called \rep{9}.

Again, the results for \SO{3} representations up to dimension $11$ are the same as already presented in \cite{Luhn:2008sa}.

\subsection{\texorpdfstring{\afive}{A5}}

The last missing subgroup of \SO{3} with an irreducible triplet representation is the icosahedral group, which is isomorphic to the alternating group on five letters \afive.
It is the largest non-abelian subgroup of \SO{3} with such a representation.
\afive can lead to golden ratio mixing when applied to neutrino model building \cite{Everett:2008et} and is, as a simple group, intrinsically anomaly-safe \cite{Chen:2015aba}.
For recent model building approaches using this group see \cite{Li:2015jxa,DiIura:2015kfa}.
Again the notation from \cite{Ludl:2009ft} is used.

The Dynkin labels of \SO{3} are split into the classes $(30\,n+2\,m)$ for $m=0,\dots,14$.
The results are shown in \autoref{tab:A5}.
\begin{table}
\centering
\tabulinesep=4pt
\begin{tabu}{>{$}l<{$}@{~~$\to$~~}>{$}c<{$}>{$}c<{$}>{$}c<{$}>{$}c<{$}>{$}c<{$}}
  \toprule
  \Lambda & \rep{1} & \rep{3} & \rep[']{3} & \rep{4} & \rep{5}\\
  \midrule
  (30\,n)   & \frac{2 n+(-1)^n+3}{4} & \frac{6 n+(-1)^{n+1}+1}{4} & \frac{6 n+(-1)^{n+1}+1}{4} & 2 n & \frac{10 n+(-1)^n-1}{4} \\
  (30\,n+2) & \frac{2 n+(-1)^{n+1}+1}{4} & \frac{6 n+(-1)^n+3}{4} & \frac{6 n+(-1)^n-1}{4} & 2 n & \frac{10 n+(-1)^{n+1}+1}{4} \\
  (30\,n+4) & \frac{2 n+(-1)^n-1}{4} & \frac{6 n+(-1)^{n+1}+1}{4} & \frac{6 n+(-1)^{n+1}+1}{4} & 2 n & \frac{10 n+(-1)^n+3}{4} \\
  (30\,n+6) & \frac{2 n+(-1)^{n+1}+1}{4} & \frac{6 n+(-1)^n-1}{4} & \frac{6 n+(-1)^n+3}{4} & 2 n+1 & \frac{10 n+(-1)^{n+1}+1}{4} \\
  (30\,n+8) & \frac{2 n+(-1)^n-1}{4} & \frac{6 n+(-1)^{n+1}+1}{4} & \frac{6 n+(-1)^{n+1}+1}{4} & 2 n+1 & \frac{10 n+(-1)^n+3}{4} \\
  (30\,n+10)& \frac{2 n+(-1)^{n+1}+1}{4} & \frac{6 n+(-1)^n+3}{4} & \frac{6 n+(-1)^n+3}{4} & 2 n & \frac{10 n+(-1)^{n+1}+5}{4} \\
  (30\,n+12)& \frac{2 n+(-1)^n+3}{4} & \frac{6 n+(-1)^{n+1}+5}{4} & \frac{6 n+(-1)^{n+1}+1}{4} & 2 n+1 & \frac{10 n+(-1)^n+3}{4} \\
  (30\,n+14)& \frac{2 n+(-1)^{n+1}+1}{4} & \frac{6 n+(-1)^n+3}{4} & \frac{6 n+(-1)^n+3}{4} & 2 n+1 & \frac{10 n+(-1)^{n+1}+5}{4} \\
  (30\,n+16)& \frac{2 n+(-1)^n-1}{4} & \frac{6 n+(-1)^{n+1}+1}{4} & \frac{6 n+(-1)^{n+1}+5}{4} & 2 n+1 & \frac{10 n+(-1)^n+7}{4} \\
  (30\,n+18)& \frac{2 n+(-1)^{n+1}+1}{4} & \frac{6 n+(-1)^n+3}{4} & \frac{6 n+(-1)^n+3}{4} & 2 (n+1) & \frac{10 n+(-1)^{n+1}+5}{4} \\
  (30\,n+20)& \frac{2 n+(-1)^n+3}{4} & \frac{6 n+(-1)^{n+1}+5}{4} & \frac{6 n+(-1)^{n+1}+5}{4} & 2 n+1 & \frac{10 n+(-1)^n+7}{4} \\
  (30\,n+22)& \frac{2 n+(-1)^{n+1}+1}{4} & \frac{6 n+(-1)^n+7}{4} & \frac{6 n+(-1)^n+3}{4} & 2 n+1 & \frac{10 n+(-1)^{n+1}+9}{4} \\
  (30\,n+24)& \frac{2 n+(-1)^n+3}{4} & \frac{6 n+(-1)^{n+1}+5}{4} & \frac{6 n+(-1)^{n+1}+5}{4} & 2 (n+1) & \frac{10 n+(-1)^n+7}{4} \\
  (30\,n+26)& \frac{2 n+(-1)^{n+1}+1}{4} & \frac{6 n+(-1)^n+3}{4} & \frac{6 n+(-1)^n+7}{4} & 2 (n+1) & \frac{10 n+(-1)^{n+1}+9}{4} \\
  (30\,n+28)& \frac{2 n+(-1)^n-1}{4} & \frac{6 n+(-1)^{n+1}+5}{4} & \frac{6 n+(-1)^{n+1}+5}{4} & 2 (n+1) & \frac{10 n+(-1)^n+11}{4} \\
  \bottomrule
\end{tabu}
\caption{Branching rules for the embedding $\afive \hookrightarrow \SO{3}$ using the triplet representation \rep{3} of \afive.
\SO{3} representations are denoted by the Dynkin labels $\Lambda$ of their highest weights.
For the conventions used, see \cite{Ludl:2009ft}.}
\label{tab:A5}
\end{table}
They show that the first singlet is contained in representation \rep{13} with Dynkin label $(12)$.

\subsection{\texorpdfstring{\dts}{Delta(27)}}

The group \dts can be embedded into \SU{3} using its triplet representation \rep{3}.
It is part of the infinite series of $\Delta(3 \cdot n^2)$ subgroups of \SU{3}.
\dts is well known in model building for the so-called geometrical spontaneous \CP violation \cite{Branco:1983tn,deMedeirosVarzielas:2011zw,Holthausen:2012dk,Fallbacher:2015rea}.
The conventions are as in \cite[Appendix~A.2]{Chen:2014tpa}.

The decomposition properties of representations of \SU{3} labelled by their Dynkin labels $(a_1,\,a_2)$ can be most easily described by splitting them into three different classes.
The Dynkin labels of these three classes take the forms $(n,\,n+3m)$, $(n,\,n+3m+1)$ and $(n,\,n+3m+2)$, where $n$ and $m$ are integers.
These classes are related to the triality classes of \SU{3} \cite{Lemire1980}.
$(n,\,n+3m)$ is in class $0$, \ie real class or class of the adjoint representation, $(n,\,n+3m+1)$ in class $2$, \ie class of the anti-fundamental representation,  and $(n,\,n+3m+2)$ in class $1$, \ie of the fundamental representation.
The resulting decomposition rules for the three classes are
\begin{subequations}
\allowdisplaybreaks
\begin{align}
  \begin{split}
    (n,\,n+3m) & \to \frac{1}{18}\, \left(f(n,m) + 16\, (-1)^n p^+(n)\right) \times \rep[_0]{1}\\
    & \qquad\qquad \oplus \frac{1}{18}\, \left(f(n,m) - 2\, q(n) \right) \times \bigoplus_{i=1}^8 \rep[_i]{1}\,,
  \end{split}\\
  (n,\,n+3m+1) & \to \frac{1}{6}\, (1 + n)\, (2 + 3 m + n)\, (3 + 3 m + 2 n) \times \crep{3}\,,\\
  (n,\,n+3m+2) & \to \frac{1}{6}\, (1 + n)\, (3 + 3 m + n)\, (4 + 3 m + 2 n) \times \rep{3}\,.
\end{align}
\end{subequations}
Hence, all real representations of \SU{3} branch to a direct sum of trivial singlets and full sets of non-trivial \dts singlets.
Moreover, the class of the fundamental \SU{3} representation yields only triplets and, accordingly, the class of the anti-fundamental only anti-triplets of \dts.
\dts is thus very much aligned with the structure of \SU{3}, making it, for example, impossible to obtain a \CP breaking representation content via spontaneous breaking.

Specialising to \SU{3} representations up to dimension $27$, the results coincide with the ones presented in \cite{Luhn:2008sa,Luhn:2011ip}.

\subsection{\texorpdfstring{\dff}{Delta(54)}}

As a second example of an \SU{3} subgroup, consider \dff embedded using its three-dimensional representation \rep[_2]{3}.
\dff is part of the $\Delta(6\cdot n^2)$ series of \SU{3} subgroups.
It turns out that, due to the additional continuous symmetries, \dff is the realised discrete symmetry group of the \dts Higgs potentials leading to geometrical \CP violation \cite{deMedeirosVarzielas:2011zw,Ivanov:2012ry,Ivanov:2012fp}.
The conventions are the same as in \cite[Appendix~A]{Fallbacher:2015rea}.

The representations of \SU{3} are again divided into the three classes described for \dts above.
The resulting decomposition rules for the three classes are
\begin{subequations}
\allowdisplaybreaks
\begin{align}
  \begin{split}
    (n,\,n+3m) & \to \frac{1}{72}\, \left[9\, (-1)^n\, \left((-1)^m\, (3 m+n+1)+n+1\right)\right.\\
    & \qquad\qquad {}+ (3 m+2 n+2)\, \left(2\, (n+1)\, (3 m+n+1)+9 (-1)^m\right)\\
    & \qquad\qquad \left.{}+32\, (-1)^n\, p^+(n)\right] \times \rep[_0]{1}\\
    & \qquad {}\oplus \frac{1}{72}\, \left[-9\, (-1)^n\, \left((-1)^m\, (3 m+n+1)+n+1\right)\right.\\
    & \qquad\qquad {}+ (3 m+2 n+2)\, \left(2\, (n+1)\, (3 m+n+1)-9 (-1)^m\right)\\
    & \qquad\qquad \left.{}+32\, (-1)^n\, p^+(n)\right] \times \rep[_1]{1}\\
    & \qquad {}\oplus \frac{1}{18}\, \left[f(n,m) - 2\, q(n) \right] \times \bigoplus_{i=1}^4 \rep[_i]{2}\,,
  \end{split}\\
  \begin{split}
    (n,\,n+3m+1) & \to \frac{1}{24}\, \left[(3 m+2 n+3)\, \left(2\, (n+1)\, (3 m+n+2)+3\, (-1)^m\right)\right.\\
    & \qquad\qquad \left.{}+3 \left((-1)^{m+1}\, (n+1)+3 m+n+2\right) (-1)^{m+n}\right] \times \crep[_2]{3}\\
    & \qquad {}\oplus \frac{1}{24}\, \left[3\, (-1)^m\, \left((-1)^{n+1}\, (3 m+n+2)-3 m-2 n-3\right)\right.\\
    & \qquad\qquad \left.{}+(n+1)\, \left(2\, (3 m+n+2)\, (3 m+2 n+3)+3\, (-1)^n\right)\right] \times \crep[_1]{3}\,,
  \end{split}\\
  \begin{split}
    (n,\,n+3m+2) & \to \frac{1}{24}\, \left[(3 m+2 n+4)\, \left(2\, (n+1)\, (3 m+n+3)-3\, (-1)^m\right)\right.\\
    & \qquad\qquad \left.{}-3\, \left((-1)^m\, (n+1)+3 m+n+3\right) (-1)^{n+m}\right] \times \rep[_2]{3}\\
    & \qquad {}\oplus \frac{1}{24}\, \left[(3 m+2 n+4)\, \left(2\, (n+1)\, (3 m+n+3)+3\, (-1)^m\right)\right.\\
    & \qquad\qquad \left.{}+3\, \left((-1)^m\, (n+1)+3 m+n+3\right) (-1)^{n+m}\right] \times \rep[_1]{3}\,.
  \end{split}
\end{align}
\end{subequations}
Although the formulas are considerably more complicated than the ones for \dts, it is easy to see that \dff is also closely aligned to the structure of \SU{3}.
Again, the real class of \SU{3} representations yields trivial singlets and complete sets of doublets (which contain the non-trivial singlets of \dts), whereas the fundamental and anti-fundamental classes contain triplets and anti-triplets, respectively.

The smallest representation of \SU{3} containing a trivial \dff singlet is the \rep{27}, which, in fact, contains three trivial \dff singlets.
This result is in agreement with \cite{Luhn:2011ip}.

\section{Conclusion}

Non-abelian finite symmetries are popular tools in model building.
They can originate from spontaneously broken continuous symmetries, thereby evading the conjectured violation of global symmetries by gravitational effects.
For model building purposes it is important to know the branching rules for this breaking. 
We have shown how to obtain these rules for the specific case of compact classical Lie groups and finite subgroups thereof using the character scalar product.
The embedding of a finite subgroup into a Lie group is specified by an explicit matrix representation of the finite group, which is then viewed as a restriction of the fundamental representation of the Lie group to the finite group.

To compute the characters of group elements for arbitrary irreducible Lie group representations, the Weyl character formula for Lie algebra characters has been reviewed and its applicability to the problem in question established.
Two different, but of course equivalent, forms of the Weyl character formula in terms of the eigenvalues of the representation matrices specifying the embedding have been presented.
These formulas have been implemented in form of the \textsc{Mathematica} package \texttt{DecomposeLGReps} that can be found \href{http://einrichtungen.ph.tum.de/T30e/codes/DecomposeLGReps}{online},%
\footnote{\url{http://einrichtungen.ph.tum.de/T30e/codes/DecomposeLGReps}}
and the usage of this package has briefly been outlined.
It can be used to compute branching rules for arbitrary non-abelian finite subgroups of the compact classical Lie groups \UU{N}, \SU{N}, \SO{N} and \USp{N}, limited only by computational power.

As an application of the package, general branching rules as functions of the Dynkin labels for various small finite groups have been derived.
The results allow to gain insight into the breaking patterns available for these finite groups.
For example, it is possible to show explicitly that the doublet representations of \tp only arise as remnants of \SU{2} spinor, in contrast to vector, representations.
Another result is that breaking \SU{3} to \dts one cannot obtain a single non-trivial singlet representation but only complete sets of non-trivial singlets.
This information is useful for (flavour) model building with spontaneously broken continuous symmetries in that certain structures of the potential can be envisaged directly from the branching rules of the symmetry.

\subsection*{Acknowledgements}

We would like to thank Michael Ratz, Patrick Vaudrevange and Andreas Trautner for valuable discussions and comments on the manuscript.
This work was supported by the research grant ``Flavor and CP in supersymmetric extensions of the Standard Model'' of Deutsche Forschungsgemeinschaft (DFG), the DFG Graduiertenkolleg 1054 ``Particle Physics at the Energy Frontier of New Phenomena'' and the TUM Graduate School.
This research was done in the context of the ERC Advanced Grant project ``FLAVOUR''~(267104).

\printbibliography

\end{document}